\documentclass[conference]{IEEEtran}

\usepackage{cite}
\usepackage{amsmath,amssymb,amsfonts}
\usepackage{graphicx}
\usepackage{textcomp}
\usepackage{xcolor}
\usepackage{booktabs}
\usepackage{multirow}
\usepackage{url}
\usepackage{balance}
\usepackage[hidelinks]{hyperref}

\newcommand{\GF}{\,GFLOPS}

\begin{document}
\bstctlcite{IEEEexample:BSTcontrol}

\title{Bandwidth, Not FLOPS: FFT Kernels,\\Matrix Units and SAR Imaging on Apple M6}

\author{\IEEEauthorblockN{Mohamed Amine Bergach}
\IEEEauthorblockA{Illumina\\
mbergach@illumina.com}}

\maketitle

\begin{abstract}
The fast Fourier transform (FFT) underlies radar, imaging and scientific computing. A classic rule
for fast GPU FFTs is to compute the largest block that fits on chip and compose larger transforms
from such blocks. We test this rule on Apple's M6 chip, whose GPU performs 29
arithmetic operations in the time it reads one byte from main memory, and which adds matrix
units to both its GPU and CPU. Comparing our kernels with Apple's MPSGraph and vDSP libraries and
MLX, we find that data movement, not arithmetic, sets FFT speed. Large batches run at
or near the main-memory bandwidth limit in every GPU library, and benchmarks that keep data in cache
overstate real throughput by up to $3.7\times$. The on-chip rule still predicts where speed
collapses: MPSGraph and MLX lose half or more of their speed once a transform outgrows the
GPU's 32\,KiB local memory. Keeping such transforms in registers avoids an extra trip through
memory: $2.2\times$ faster than MPSGraph, and $4.4\times$ with half-precision storage. The
GPU's matrix units do not help, because recasting the FFT as matrix products adds as much arithmetic
as they save. The CPU's matrix unit, 47 times faster than its vector units, does: our kernel for it
beats vDSP by up to $5.3\times$. End to end, a $4096\times4096$ synthetic aperture radar image takes
8.1\,ms on the GPU (6.0\,ms with half-precision intermediates), $14$--$19\times$ faster than a CPU
reference. Running the CPU's matrix unit alongside the GPU gains nothing: both share one memory
bandwidth.
\end{abstract}


\section{Introduction}

The fast Fourier transform (FFT) is a workhorse of radar, imaging and signal processing. A 2015 study
of FFTs on chips that combine a CPU and a GPU~\cite{bergach2015thesis,bergach2014scaling,bergach2015conference}
reduced fast GPU FFT design to one rule: compute the largest FFT block that fits in the GPU's fast
on-chip memory, and build larger transforms from such blocks. On Apple's M1 the rule produced a kernel
that beat Apple's own vDSP library~\cite{bergach2026fft}, a radar imaging pipeline that was fastest
when its stages were fused inside one block~\cite{bergach2026sar}, and a safe way to use half precision
(FP16) for radar~\cite{bergach2026fp16}.

Apple's M6 changes the ground under that rule in three ways. First, its GPU computes far faster than
it can read main memory: 29 floating-point operations per byte, almost eight times what a 4096-point
FFT performs per byte (Section~\ref{sec:hier}). Second, a large system-level cache (SLC) now sits
between the GPU and main memory, so the same FFT runs at different speeds depending on where its data
happens to live. Third, both the GPU and the CPU now include matrix units, which offer much more
arithmetic, but only for matrix products (Table~\ref{tab:device}). This paper measures what these
changes mean for FFT performance, from single kernels to a complete radar imaging pipeline. Our
contributions are:

\begin{enumerate}
\item \emph{Memory, not arithmetic, sets the speed.} Large batches of every GPU FFT library we tested
  run at or near the main-memory bandwidth limit, and benchmarks that leave data in cache overstate
  real throughput by up to $3.7\times$ (Sections~\ref{sec:hier} and~\ref{sec:landscape}).
\item \emph{The on-chip rule still predicts the cliffs.} Apple's MPSGraph and the MLX framework both
  lose half or more of their speed exactly when a transform stops fitting in the GPU's 32\,KiB on-chip
  memory (Section~\ref{sec:landscape}).
\item \emph{Larger transforms in a single pass.} New kernels keep 8192- and 16384-point transforms in
  registers, so their data crosses main memory only once: $2.2\times$ and $1.85\times$ faster than
  MPSGraph, and $4.4\times$ with FP16 storage (Sections~\ref{sec:onepass} and~\ref{sec:fp16}).
\item \emph{When matrix units help.} A simple bound predicts whether a matrix unit can speed up an FFT.
  The GPU's matrix units cannot. The CPU's can: our kernel for it beats vDSP by $1.49\times$ on
  batches of 4096-point transforms and by up to $5.3\times$ on the columns of a matrix
  (Sections~\ref{sec:mma} and~\ref{sec:sme}).
\item \emph{A complete radar pipeline.} A validated $4096\times4096$ synthetic aperture radar (SAR)
  image takes 8.1\,ms on the GPU, or 6.0\,ms with FP16 intermediates, $14$--$19\times$ faster than a
  CPU reference. Running the CPU's matrix unit alongside the GPU adds nothing, because both share the
  same memory bandwidth (Section~\ref{sec:sar}).
\end{enumerate}
All kernels, benchmark programs, raw logs and plotting scripts are available
online~\cite{applesiliconfft}.

\section{Related Work}\label{sec:related}

\textit{GPU FFTs.} Fast GPU FFTs have long been built from blocks computed in on-chip shared memory,
starting with early CUDA work~\cite{govindaraju2008fft}. NVIDIA's cuFFT~\cite{nvidia_cufft}, the
cross-platform VkFFT, which also runs on Metal~\cite{tolmachev2023vkfft}, and TurboFFT, which adds
fault tolerance~\cite{wu2025turbofft}, all follow this structure. cuFFTDx~\cite{nvidia_cufftdx} lets
an FFT be called from inside a user's own kernel; our radar pipeline uses the same idea to fuse
stages. The on-chip rule was derived on Intel integrated
GPUs~\cite{bergach2014scaling,bergach2015conference,bergach2015thesis} and carried over to Apple's
M1~\cite{bergach2026fft,bergach2026sar,bergach2026fp16}. On the CPU side, FFTW is the reference
auto-tuned library~\cite{frigo2005fftw}, and butterfly arithmetic and SIMD instruction scheduling are
studied in~\cite{bergach2026butterfly,bergach2026shortest}.

\textit{Memory-bound performance.} Our speed limits form a roofline model~\cite{williams2009roofline}
with one roof per memory level. The number of passes over slow memory, which drives most of our
results, is the quantity that the I/O complexity of the FFT bounds~\cite{hong1981io}.

\textit{Matrix units.} tcFFT runs half-precision FFTs on NVIDIA's Tensor Cores~\cite{li2021tcfft}.
Apple's earlier CPU matrix unit (AMX) was documented by reverse engineering~\cite{corsix_amx}. From
M4 on, the CPU implements Arm's Scalable Matrix Extension (SME)~\cite{arm_sme}, for which generated
matrix-multiplication kernels exceed 2.3\,TFLOPS on M4~\cite{remke2024hellosme}. We use this unit for
the FFT, with the constant-geometry ordering that Pease designed for parallel
processors~\cite{pease1968adaptation,vanloan1992fft}.

\textit{Apple Silicon.} Hübner et~al.\ evaluate the M1 to M4 chips for high-performance
computing~\cite{hubner2025apple}. The lever we use for FP16 storage, moving fewer bytes per value, also
makes a 4-bit key-value cache faster than an FP16 one in language-model inference on Apple
Silicon~\cite{bergach2026int4}.

\section{Platform and Methodology}\label{sec:platform}

\begin{table}[t]
\centering
\caption{Test Machine (\texttt{sysctl} and Metal API; Bandwidths and Peak Rates Measured,
Section~\ref{sec:hier})}
\label{tab:device}
\footnotesize
\begin{tabular}{@{}lp{0.70\columnwidth}@{}}
\toprule
SoC & Apple M6, macOS 27.0 (26A428), AC power \\
GPU & 12 cores, Metal 4, family \texttt{apple10}; 32\,KiB threadgroup memory, up to 1024 threads per
threadgroup \\
GPU peak & 4.38\,TFLOPS FP32 FMA, 8.46\,TFLOPS FP16 FMA \\
CPU & 12 cores in three tiers: 2 Super, 4 Performance, 6 Efficiency \\
CPU caches & L1d 128 / 64 / 96\,KiB; L2 20 / 20 / 8\,MiB, one per Super / Performance /
Efficiency cluster \\
Matrix units & CPU: Arm SME/SME2~\cite{arm_sme}, 512-bit vectors; GPU: Metal~4 tensor ops \\
Memory & 32\,GiB unified; 150\,GB/s GPU streaming bandwidth \\
\bottomrule
\end{tabular}
\end{table}

Table~\ref{tab:device} lists the test machine. In Apple's Metal API, GPU threads run in
\emph{threadgroups}; the threads of one threadgroup share 32\,KiB of fast on-chip \emph{threadgroup
memory}, the ``local memory'' of the on-chip rule.

\textit{Batch sizes.} Speed depends strongly on where the data lives, so we size each batch to fill one
level of the memory hierarchy. The \emph{footprint} of a batch is its input plus its output: $2BN\cdot8$
bytes for $B$ complex FP32 transforms of $N$ points. We use footprints of 1\,MiB (fits the GPU's L2
cache), 8\,MiB (fits the SLC) and 256\,MiB (main memory, DRAM), with $B\ge1$.

\textit{Hot and cold runs.} A \emph{hot} measurement repeats the FFT back to back on the same buffers,
so data can stay in cache between repetitions; each sample moves at least 64\,MiB, and we report the
time per FFT. A \emph{cold} measurement first writes a 512\,MiB buffer, which evicts both caches, and
then times a single FFT. GPU times come from the GPU's own timestamps. Unless stated otherwise we report
the median of 11--15 interleaved samples; the median spread (interquartile range) is 2--4\% of the
median. The CPU scaling measurements (Table~\ref{tab:cpu}) use one 2-s window per cell.

\textit{Throughput.} We count $5N\log_2N$ floating-point operations per complex FFT, the usual
convention~\cite{bergach2026fft}. Because the limits we find are bandwidth limits, we also report
effective bandwidth: bytes in plus bytes out, divided by time.

\textit{Correctness.} Every GPU result is checked against vDSP (or NumPy for MLX) on the first
$\min(B,4)$ transforms; all FP32 results agree to a relative error of $8.2\times10^{-7}$ or better.
For FP16 storage we compare with vDSP on the same FP16-rounded input and report the
signal-to-quantization-noise ratio, SQNR $=-20\log_{10}(\text{relative error})$. Inputs are random and
uniform on $[-1,1]$.

\textit{Baselines.} MPSGraph is Apple's GPU graph library (\texttt{fastFourierTransform}~\cite{apple_mpsgraph});
we time it on the GPU, like our kernels. MLX 0.29.3 is Apple's machine-learning framework
(\texttt{mx.fft.fft}~\cite{mlx2023}); it exposes no GPU timestamps, so we time it by wall clock, and
small sizes include its per-call overhead. vDSP is Apple's CPU signal-processing library
(\texttt{vDSP\_fft\_zop}~\cite{apple_vdsp_fft}); we run it on one thread at the highest priority, which
the scheduler places on a Super core.

\section{The M6 Memory Hierarchy}\label{sec:hier}

\begin{table}[t]
\centering
\caption{GPU Bandwidth vs.\ Working Set per Buffer (Hot, Median of 9; Copy Counts Bytes Read Plus
Written)}
\label{tab:bw}
\footnotesize
\begin{tabular}{@{}rrrl@{}}
\toprule
\textbf{Working set} & \textbf{Copy GB/s} & \textbf{Read GB/s} & \textbf{Level} \\
\midrule
256\,KiB & 423 & 251 & L2 (under-occupied) \\
1\,MiB & 671 & 534 & L2 \\
2\,MiB & 311 & 639 & L2 / SLC boundary \\
4--8\,MiB & 264--317 & 304--308 & SLC \\
16\,MiB--1\,GiB & 145--153 & 150--155 & DRAM \\
\bottomrule
\end{tabular}
\end{table}

The GPU sees three speeds of memory (Table~\ref{tab:bw}): about 530--670\,GB/s while the data fits its
L2 cache (1--2\,MiB), about 300\,GB/s while it fits the SLC (up to a total of about 16\,MiB), and
\textbf{150\,GB/s} from main memory, flat up to 1\,GiB. With a measured peak of 4.38\,TFLOPS (FP32),
the GPU can perform 29 operations for every byte it reads from main memory.

An FFT that reads its input once and writes its output once, a single \emph{pass}, moves 16 bytes per
point (FP32 complex, in and out) and performs $5\log_2N$ operations per point. From main memory its
speed is therefore capped at
\begin{equation}
  P_\text{DRAM}(N) = 150\,\text{GB/s}\times\frac{5\log_2 N}{16\,\text{B}},
  \label{eq:ceiling}
\end{equation}
that is, 562\GF{} at $N=4096$ and 609\GF{} at $N=8192$. We call this the main-memory limit (the
dashed ``ceiling'' lines in the figures). Each extra pass through main memory divides it, and storing
values in FP16 (8 bytes per point) doubles it.

\subsection{Hot vs.\ Cold: Cached Benchmarks Overstate Speed}\label{sec:hotcold}

\begin{table}[t]
\centering
\caption{One 4096-Point Kernel (512 Threads per Threadgroup) vs.\ Batch Size. Hot: Data Left in
Cache by the Previous Run; Cold: Caches Flushed First}
\label{tab:batch}
\footnotesize
\begin{tabular}{@{}rrrrr@{}}
\toprule
\textbf{Batch} & \textbf{In+out MiB} & \textbf{Hot GFLOPS} & \textbf{Hot GB/s} & \textbf{Cold GFLOPS} \\
\midrule
12 & 0.75 & 1193 & 318 & 342 \\
24 & 1.5 & \textbf{1534} & 409 & 401 \\
256 & 16 & 1064 & 284 & 385 \\
384 & 24 & 605 & 161 & 430 \\
1024 & 64 & 585 & 156 & 507 \\
16384 & 1024 & 565 & 151 & 558 \\
\bottomrule
\end{tabular}
\end{table}

Table~\ref{tab:batch} runs one 4096-point kernel on batches of growing size. With hot data, speed peaks
at 1534\GF{} (35\% of the GPU's peak) for a batch of 24, stays near 1.05\,TFLOPS while the batch fits
the SLC, and falls to the main-memory limit once the batch exceeds about 16\,MiB. Cold runs rise toward
the same limit, with a dip between 3 and 16\,MiB (448 to 385\GF{}); small cold batches are dominated by
launch latency and first-touch cache misses. The best hot number is $2.7\times$ what a streaming
pipeline would see at this size, and the gap reaches $3.7\times$ at $N=128$ (Section~\ref{sec:landscape}).
A hot small-batch benchmark on M6 therefore measures the cache, not the FFT. We report both, with the
footprint, throughout.

A batch of 24 (two threadgroups per GPU core) runs 29\% faster per FFT than a batch of 12 (one per
core), which suggests that two 32\,KiB threadgroups fit on one core. The M1-era slowdown at about 128
registers per thread is also no longer visible through the API
(\texttt{maxTotalThreadsPerThreadgroup} stays at 1024).

\subsection{CPU: vDSP Runs on a Shared Matrix Unit}\label{sec:cpu}

\begin{table}[t]
\centering
\caption{CPU Throughput (Total GFLOPS) vs.\ Number of Threads. UI: Highest Priority (Super and
Performance Cores); BG: Background Priority (Efficiency Cores)}
\label{tab:cpu}
\footnotesize
\setlength{\tabcolsep}{3.5pt}
\begin{tabular}{@{}lrrrrrrr@{}}
\toprule
 & \multicolumn{5}{c}{\textbf{UI threads}} & \multicolumn{2}{c}{\textbf{BG threads}} \\
\cmidrule(lr){2-6}\cmidrule(l){7-8}
\textbf{Workload} & 1 & 2 & 4 & 6 & 12 & 1 & 6 \\
\midrule
NEON FMA loop & 49 & 96 & 186 & 278 & 447 & 12 & 60 \\
\texttt{sgemm} $512^3$ & 1659 & 1906 & 1908 & 1910 & 2253 & 127 & 123 \\
vDSP FFT 4096 & 203 & 210 & 204 & 197 & 246 & 19 & 19 \\
\bottomrule
\end{tabular}
\end{table}

Table~\ref{tab:cpu} shows how Apple's CPU libraries use the hardware. A plain loop on the CPU's vector
units (NEON) scales linearly with the number of threads, as expected. Matrix multiplication
(\texttt{cblas\_sgemm}), which Apple's Accelerate framework runs on the CPU's matrix unit, does not: it
is flat from 1 to 6 threads. Neither does vDSP's FFT. The threads do spread over all six Super and
Performance cores, so these six cores must share one matrix unit, with a separate, much slower unit on
the Efficiency cores. Three consequences follow. Running vDSP on more threads is not a stronger
baseline. vDSP's FFT uses only about 11\% of the matrix unit's multiplication rate. And the vector units
of the other eleven cores sit idle while it runs. (The vector loop is limited by latency, with 8
accumulation chains, so 49\GF{} per core is a lower bound on the vector peak.)

\section{The FFT Landscape on M6}\label{sec:landscape}

\begin{figure*}[t]
\centering
\includegraphics[width=\textwidth]{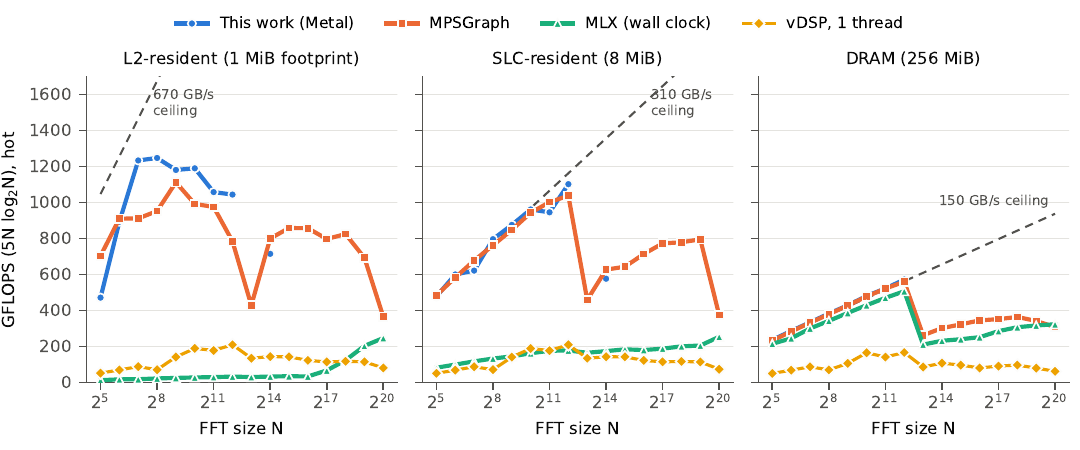}
\caption{Hot throughput vs.\ transform size $N$ for each memory level. ``This work'' is our best
existing kernel for each $N$ (none for $N=8192$ or $N>16384$ in this figure; see
Section~\ref{sec:onepass}). Dashed: single-pass ceiling for the level's measured bandwidth,
Eq.~(\ref{eq:ceiling}). The cold-cache version is in the repository~\cite{applesiliconfft}.}
\label{fig:landscape}
\end{figure*}

Fig.~\ref{fig:landscape} shows hot throughput for every library, transform size and memory level. The
main observations follow.

\textit{In main memory, every GPU library hits the same wall.} For $N\le4096$, our kernels and MPSGraph
run at 149--154\,GB/s, within 3\% of the 150\,GB/s limit, and MLX runs 8--13\% lower (131--138\,GB/s).
In operations, this is the curve of Eq.~(\ref{eq:ceiling}), from 234\GF{} at $N=32$ to 562\GF{} at
$N=4096$. No FFT kernel can go faster at these sizes; only moving fewer bytes can help.

\textit{The on-chip rule predicts where libraries slow down.} At $N=8192$ an FP32 transform (64\,KiB) no
longer fits in 32\,KiB of threadgroup memory, and both third-party libraries fall off. MPSGraph halves,
from 560 to 262\GF{} (64.6\,GB/s effective), and stays at 65--69\,GB/s up to $2^{18}$, which means two
passes through main memory (58 and 49\,GB/s at $2^{19}$ and $2^{20}$). MLX falls from 507 to 211\GF{}
and stays at 50--55\,GB/s, about three passes. A rule derived in 2015 on Intel integrated GPUs thus
predicts exactly where the cliff is, on a 2026 chip and in code we did not write. (Our earlier kernels
include no single-kernel $N=8192$ transform; Section~\ref{sec:onepass} adds one.)

\textit{In cache, our kernels are faster.} With a 1\,MiB footprint, our kernels beat MPSGraph by
$1.06$--$1.35\times$ for $N=128$--$4096$ (for example 1247 vs.\ 952\GF{} at $N=256$). MPSGraph wins at
$N=32$ (703 vs.\ 471), where our kernel runs only 8 threads per threadgroup, and at $N=16384$. With an
8\,MiB footprint (SLC) the two are within 8\%, both near the SLC limit.

\textit{Small cold batches are dominated by launch cost.} At 1\,MiB cold ($N\le2^{16}$), our kernels and
MPSGraph drop to 176--443\GF{}, and MLX to 8--26\GF{}. Pipelines built from many small FFT launches pay
this cost at every stage, which argues for fusing stages~\cite{bergach2026sar}, whatever the bandwidth.

\textit{The CPU and MLX.} Single-threaded vDSP reaches between 50 and 210\GF{} (highest at $N=4096$);
the GPU is $2.8$--$5.5\times$ faster on main-memory batches. MLX, called from Python, costs about
110--160\,$\mu$s per call at a 1\,MiB footprint, where the GPU work itself takes a few microseconds. It is
competitive only for large batches and for $N\ge2^{18}$, so we treat it as a user-level rather than a
kernel-level number.

\section{One Pass for $N=8192$ and $16384$}\label{sec:onepass}

From $N=8192$ on, every library in Section~\ref{sec:landscape} makes at least two passes through main
memory and loses half its speed. The usual remedy when a transform does not fit on chip is a two-pass
(``four-step'') FFT. On M6 there is a better option: the registers of one threadgroup hold far more
data than its 32\,KiB of threadgroup memory. The whole transform can stay in registers, as long as the
data exchanged between FFT stages can pass through a buffer smaller than the transform.

\textit{Kernel.} One threadgroup computes one transform: 512 threads each hold 16 complex values for
$N=8192$, and 1024 threads each hold 16 for $N=16384$. The kernel is a mixed-radix Stockham FFT, with
stages of radix 16, 8, 8, 8 and 16, 16, 8, 8, built from the radix-8 and radix-16 butterflies
of~\cite{bergach2026fft}. Between stages, the threads exchange data through the 32\,KiB buffer (8192
floats) one real component at a time, and for $N=16384$ in two halves. Every point is read from main
memory once and written once, and all arithmetic and exchanges are FP32. With two halves, a register
may be refilled in the first half while its old value is still due to be written in the second, so reads
go to a second register array; a regression test forces $N=8192$ through this path.

\begin{table}[t]
\centering
\caption{Batches in Main Memory (256\,MiB FP32 Footprint), Hot / Cold GFLOPS. Limit From
Eq.~(\ref{eq:ceiling})}
\label{tab:onepass}
\footnotesize
\setlength{\tabcolsep}{4pt}
\begin{tabular}{@{}lrrrr@{}}
\toprule
\textbf{$N$ (batch)} & \textbf{One pass} & \textbf{MPSGraph} & \textbf{Four-step} & \textbf{Limit} \\
\midrule
8192 (2048) & \textbf{601 / 584} & 274 / 270 & -- & 609 \\
16384 (1024) & \textbf{570 / 565} & 308 / 307 & 311 / 307 & 656 \\
\bottomrule
\end{tabular}
\end{table}

\textit{Results.} At $N=8192$ the one-pass kernel reaches 99\% of the main-memory limit, $2.2\times$
MPSGraph (Table~\ref{tab:onepass}). At $N=16384$ it reaches 87\% of the limit: $1.85\times$ MPSGraph
and $1.8\times$ our own two-pass kernel. The relative errors are $3.6\times10^{-7}$ and
$4.8\times10^{-7}$.

\textit{When not to use it.} With one threadgroup per transform, a batch needs enough transforms to keep
the 12 GPU cores busy. With a 1\,MiB footprint ($B=4$), the $N=16384$ kernel reaches only 221\GF{},
against 815 for MPSGraph. At $B=32$ it still trails (486 vs.\ 666): with one 1024-thread threadgroup per
core, 32 transforms fill fewer than three rounds of 12 cores. At $N=8192$, with 512-thread threadgroups,
one pass already wins at $B=8$ (740 vs.\ 466). \emph{The choice therefore depends on $N$: one pass at
$N=8192$ for every batch we measured, and at $N=16384$ only for large, main-memory batches; for smaller
batches, split each transform across threadgroups.}

\section{FP16 Storage}\label{sec:fp16}

\begin{figure}[t]
\centering
\includegraphics[width=\columnwidth]{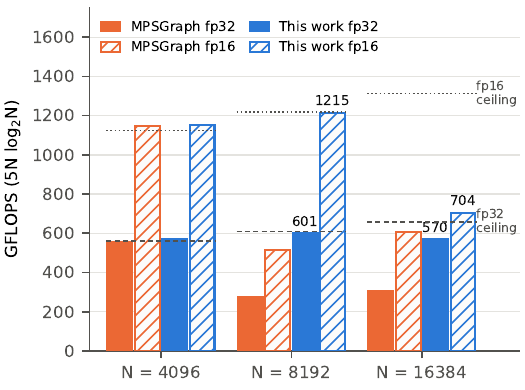}
\caption{Batches in main memory, hot, with the same number of FFTs in FP32 and FP16. Solid: FP32
storage; hatched: FP16 storage with FP32 arithmetic. Dashed and dotted lines: FP32 and FP16
single-pass ceilings.}
\label{fig:onepass}
\end{figure}

\begin{table}[t]
\centering
\caption{FP16 Storage With FP32 Arithmetic, Batches in Main Memory, Hot GFLOPS. SQNR Against vDSP on
the Same FP16-Rounded Input}
\label{tab:fp16}
\footnotesize
\setlength{\tabcolsep}{3pt}
\begin{tabular}{@{}rrrrrr@{}}
\toprule
 & \multicolumn{2}{c}{\textbf{FP16 storage}} & \textbf{FP32} & \textbf{FP16} & \textbf{SQNR (dB)} \\
\cmidrule(lr){2-3}
\textbf{$N$} & \textbf{Ours} & \textbf{MPSGraph} & \textbf{MPSGraph} & \textbf{limit} & \textbf{ours / MPSG.} \\
\midrule
4096 & 1152 & 1146 & 558 & 1125 & 67.7 / 73.7 \\
8192 & \textbf{1215} & 513 & 274 & 1219 & \textbf{73.7} / 70.6 \\
16384 & 704 & 609 & 308 & 1313 & 73.6 / 70.7 \\
\bottomrule
\end{tabular}
\end{table}

Once an FFT runs at the main-memory limit, only moving fewer bytes can make it faster. Storing inputs
and outputs in half precision (FP16) while computing in FP32 halves the traffic (Fig.~\ref{fig:onepass},
Table~\ref{tab:fp16}):
\begin{itemize}
\item At $N=4096$, both our kernel and MPSGraph reach the doubled limit (about 1150\GF{}, $2.0\times$
  FP32).
\item The one-pass kernel reaches 1215\GF{} at $N=8192$, 99.7\% of its limit: $4.4\times$ MPSGraph
  in FP32 and $2.4\times$ MPSGraph in FP16, which still makes two passes.
\item At $N=16384$, halving the bytes exposes the kernel's own limit: 704\GF{} is only 80\,GB/s. A
  variant with 512 threads holding 32 values each is slower (453\GF{}), so the limit is hiding latency
  with a single 1024-thread threadgroup per core, not bandwidth.
\end{itemize}

Accuracy depends on where FP16 is used. Our one-pass kernels keep the exchange between stages in FP32
and reach 73.6--73.7\,dB SQNR, about 3\,dB better than MPSGraph in FP16 at the same sizes. Our
$N=4096$ kernel also stores that exchange in FP16 and is 6\,dB \emph{worse} than MPSGraph (67.7 vs.\
73.7\,dB). FFT outputs grow with $\sqrt N$ for noise-like data and up to $N$ for coherent data, so
storing real radar data in FP16 needs block-floating-point scaling~\cite{bergach2026fp16}; the inputs
here are uniform noise. The fastest FFT we measured on M6 is the FP16-storage 4096-point transform on a
batch held in the SLC: 1846\GF{}, 42\% of the GPU's FP32 peak.

\section{The GPU Matrix Units}\label{sec:mma}

M5-class and later GPUs add matrix hardware to each core, reached from Metal~4 through tensor
operations (\texttt{mpp::\allowbreak tensor\_ops::\allowbreak matmul2d}). We call them from our own
shaders, compiled at run time with version 4.0 of the Metal Shading Language.

\begin{table}[t]
\centering
\caption{GPU Matrix Units. Top: Matrix Multiplication Through \texttt{matmul2d} (Square Size,
$64\times64$ Tiles, 4 SIMD Groups, Median of 7). Bottom: 4096-Point FFT With FP16 Storage, Hot GFLOPS;
``Matrix Products Only'' Runs the Three Products Without Twiddle Factors}
\label{tab:mma}
\footnotesize
\begin{tabular}{@{}lrrr@{}}
\toprule
\textbf{Matrix multiplication} & \textbf{1024} & \textbf{2048} & \textbf{4096} \\
\midrule
half$\times$half$\to$float & 19.6\,TF & 17.3\,TF & 11.0\,TF \\
float$\times$float$\to$float & 4.7\,TF & 4.8\,TF & 4.3\,TF \\
\bottomrule
\end{tabular}

\medskip
\begin{tabular}{@{}lrrr@{}}
\toprule
\textbf{4096-point FFT, FP16 storage} & \textbf{L2} & \textbf{SLC} & \textbf{DRAM} \\
\midrule
Matrix-unit FFT (radix 16) & 520 & 932 & 1034 \\
\quad matrix products only & 948 & 1377 & 1129 \\
Ordinary radix-8 kernel & \textbf{1294} & \textbf{1792} & \textbf{1119} \\
\bottomrule
\end{tabular}
\end{table}

\textit{Fast matrix products.} In FP16 with FP32 accumulation, the units multiply matrices at
17--19.6\,TFLOPS for sizes 1024--2048 (Table~\ref{tab:mma}): $2.3\times$ the GPU's FP16 peak and
$4.5\times$ its FP32 peak. In FP32 they reach 4.3--4.8\,TFLOPS, about the ordinary FP32 rate, so they
help only in FP16. Unlike on M1, the matrix path is genuinely faster than the ordinary units, in FP16.

\textit{An FFT built from matrix products.} Each stage of a radix-16, 4096-point FFT can be written as
one matrix product: a fixed $32\times32$ real matrix (the 16-point DFT, with real and imaginary parts
separated) times a $32\times256$ matrix of data in threadgroup memory. Between stages, each thread
applies the twiddle factors to one column and moves it to where the next stage expects it. The result is
correct (66.1\,dB SQNR, against 67.7\,dB for our ordinary FP16 kernel) but slow: $0.40$--$0.52\times$
the speed of the ordinary kernel when the data is in cache, and $0.92\times$ from main memory
(Table~\ref{tab:mma}); its L2 speed also varies widely (interquartile range 362--596\GF{}). In cache,
the matrix products alone, without twiddle factors, are already slower than the complete ordinary FFT.
They run at about 6\,TFLOPS, not 19.6, because the matrices are small ($K=32$), their operands sit in
threadgroup memory, and only one 256-thread threadgroup runs per core.

\textit{Why the matrix units cannot pay.} Written as a dense matrix product, a radix-$r$ stage costs
$8r$ operations per point, while the FFT needs only $5\log_2r$. The matrix form therefore does
$3.2\times$ more arithmetic at $r=4$, $4.3\times$ at $r=8$, $6.4\times$ at $r=16$ and $17\times$ at
$r=64$. The matrix units are only $4.5\times$ faster than the ordinary FP32 units our FFT kernels use.
Even at peak speed, the matrix route only breaks even at $r=8$ and loses for $r\ge16$; at $r=4$ the
products ($K=8$) are too small to run anywhere near peak. From main memory, every FFT is limited by
bandwidth anyway. The M1 verdict thus stands on M6 for a new reason: on M1 the matrix path lost to the
cost of moving data into place; on M6 it loses to the extra arithmetic.

\section{The CPU Matrix Unit (SME2)}\label{sec:sme}

The M6 CPU also has a matrix unit, programmed through Arm's SME2 instructions. Its vectors are 512 bits
wide, so its accumulator array, ZA, holds four $16\times16$ FP32 tiles. We program it directly with
compiler intrinsics. The code must be built with \texttt{-mcpu=apple-m4}: a generic \texttt{armv9}
target lets the compiler emit ordinary (non-streaming) SVE instructions, which these cores do not have.

\textit{The unit.} Its basic instruction, \texttt{FMOPA}, adds a $16\times16$ outer product (512
operations) to a tile, with a latency of 0.91\,ns. With four tiles in flight, one thread reaches
2.29\,TFLOPS, consistent with the 2.3\,TFLOPS reported for M4~\cite{remke2024hellosme}. Two, four or six
threads on the Super and Performance cores reach the same 2.2--2.3\,TFLOPS: one thread saturates the
unit that Section~\ref{sec:cpu} found to be shared. The unit reads from the L2 cache at 286\,GB/s into
vector registers (466\,GB/s directly into ZA) and writes tile rows at 286\,GB/s. At full speed that is
about one 64-byte load and one 64-byte store per instruction, so for an FFT stage, which writes back
everything it computes, the stores set the speed.

\textit{A shared resource.} Our first measurements coincided with heavy activity of the system's audio
and conferencing services (\texttt{coreaudiod} at 56\% CPU, \texttt{avconferenced} at 32\%). During it,
the unit's single-thread rate fell to 1.40\,TFLOPS and vDSP's 4096-point FFT to 115\GF{}, against
2.29\,TFLOPS and 206\GF{} otherwise. We did not identify which process used the unit; the same services
at their usual low activity did not affect it. All numbers below come from an idle unit, checked before
each run, with the busiest processes logged. On the CPU, the matrix unit is a system-wide shared resource
and needs the same care as the GPU's cache state (Section~\ref{sec:hotcold}).

\textit{Why the matrix route pays here.} The bound of Section~\ref{sec:mma} compares a matrix unit with
the vector units an ordinary FFT would use. On the GPU that ratio is $4.5\times$: about the $4.3\times$
extra arithmetic of radix 8, and below the $6.4\times$ of radix 16. On the CPU it is
2.29\,TFLOPS\,/\,49\GF{} $\approx47\times$ against one core's vector units, and still about $11\times$
against vDSP. The extra arithmetic is affordable here.

\textit{Kernel.} A radix-8 butterfly, twiddle factors included, is one real $16\times16$ matrix acting on
16 real inputs (8 complex values),
\begin{equation}
  M=\mathrm{diag}(w_\mathrm{post})\,F_8\,\mathrm{diag}(w_\mathrm{pre}),
\end{equation}
where $F_8$ is the 8-point DFT and the diagonal matrices hold the twiddle factors. We apply it as 16
\texttt{FMOPA} updates into a tile whose 16 columns are 16 independent transforms (``batch in lanes''),
with four tiles, that is four butterflies, in flight. The data layout interleaves the 16 transforms:
point $n$ of all 16 is one 64-byte vector of real parts followed by one of imaginary parts. The usual
Stockham ordering writes each butterfly's outputs far apart, and scattered stores are expensive: one
stage takes 25\,ns per four butterflies when the outputs are contiguous, but 68\,ns and 84\,ns when they
are 8 and 64 positions apart, against 15\,ns for the arithmetic alone. We therefore use Pease's
constant-geometry ordering~\cite{pease1968adaptation}, in which every stage reads $x[J+rN/8]$ and writes
$y[8J+q]$, so the four tiles always store 4\,KiB contiguously. The output comes out in base-8
digit-reversed order, which the routines that write the final layout undo at no cost, because they write
rows at arbitrary addresses anyway.

\begin{table}[t]
\centering
\caption{CPU Matrix-Unit (SME2) FFT vs.\ vDSP, One Thread, FP32, GFLOPS. Relative Error vs.\ vDSP
$\le2.5\times10^{-7}$}
\label{tab:sme}
\footnotesize
(a) Batches of transforms: hot, 16 transforms in L2; cold, 256\,MiB. Our kernel in its own layout,
with Stockham ordering, and including conversion from and to vDSP's layout.\\[2pt]
\begin{tabular}{@{}rlrrrr@{}}
\toprule
\textbf{$N$} & & \textbf{SME} & \textbf{Stockham} & \textbf{+conv.} & \textbf{vDSP} \\
\midrule
64 & hot & 252 & 243 & 126 & 72 \\
512 & hot & 255 & 180 & 127 & 142 \\
4096 & hot & \textbf{307} & 161 & 127 & 206 \\
4096 & cold & 198 & 123 & 83 & 164 \\
32768 & hot & 243 & 123 & 102 & 140 \\
32768 & cold & 187 & 104 & 79 & 113 \\
\bottomrule
\end{tabular}

\medskip
(b) Transforms along the columns of an $N\times W$ matrix (separate real and imaginary planes); vDSP:
the better of strided transforms and transpose, transform, transpose back.\\[2pt]
\begin{tabular}{@{}rrrrr@{}}
\toprule
\textbf{$N$} & \textbf{$W$} & \textbf{SME} & \textbf{vDSP} & \textbf{Speedup} \\
\midrule
512 & 256 & 263 & 50 & 5.3$\times$ \\
512 & 65536 & 97 & 26 & 3.7$\times$ \\
4096 & 256 & 216 & 56 & 3.9$\times$ \\
4096 & 8192 & 115 & 43 & 2.7$\times$ \\
32768 & 1024 & 82 & 30 & 2.7$\times$ \\
\bottomrule
\end{tabular}
\end{table}

\textit{Batches of transforms.} In its own layout, the kernel beats vDSP at every size
(Table~\ref{tab:sme}a): $1.49\times$ at $N=4096$ in cache (307 vs.\ 206\GF{}), $1.21\times$ from main
memory, and $1.5$--$3.5\times$ at the other sizes. The constant-geometry ordering is worth up to
$2\times$ over Stockham. Converting from and to vDSP's layout (one array per signal) costs more than the
FFT itself, however, so as a drop-in replacement for vDSP the kernel loses ($0.5$--$0.9\times$) except at
$N=64$.

\textit{Columns of a matrix need no conversion.} With 16 transforms side by side, 16 adjacent columns of a
row-major matrix are one contiguous vector per row, so transforms along the columns can read and write
the matrix directly. Such column FFTs are the azimuth FFT of radar imaging and the second pass of a 2-D
FFT. Processing 64 columns at once (four tiles, one butterfly across 64 columns: 256\,B per row access
and one shared matrix) nearly doubles the speed from main memory compared with 16 columns. For the
shapes in Table~\ref{tab:sme}b, the kernel is $2.7$--$5.3\times$ faster than the better of vDSP's two
options. Over all twelve shapes we measured, the range is $1.5$--$5.3\times$; the smallest gain is for 16
columns at $N=32768$, where vDSP's transpose route is cheap. At $N=32768$ the 64-column working set
(16\,MiB, double-buffered) exceeds the L2 cache, and a two-level split would be needed.

\textit{Threads.} Like vDSP (Section~\ref{sec:cpu}), the kernel does not speed up with more threads:
303, 299, 276, 280 and 286\GF{} at 1, 2, 3, 4 and 6 threads for $N=4096$ in cache (vDSP: 198--215).
From main memory, a second thread helps a little (183 to 218\GF{}), probably through more outstanding
cache misses. The SME FFT is a one-core engine that leaves the other eleven cores free. Separately,
vDSP's strided multi-transform call (\texttt{vDSP\_fftm\_zop}) collapses from 25--47 to 1--7\GF{} in total
when two or more threads run it at once, even with one setup per thread; we did not investigate this
further.

\section{End to End: Range--Doppler SAR}\label{sec:sar}

Synthetic aperture radar (SAR) forms an image from radar echoes collected along a flight path. Its
classic range--Doppler algorithm~\cite{cumming2005sar} is a chain of FFTs along both image axes, so it
tests whether our kernel-level conclusions hold in a complete application. We process a
$4096\times4096$ scene (pulses $\times$ range samples; separate real and imaginary FP32 planes,
128\,MiB per complex image). The scene is simulated in an L-band stripmap geometry
($\lambda=0.24$\,m, $B=100$\,MHz, $f_s=120$\,MHz, $v=200$\,m/s, PRF 500\,Hz, 8\,m antenna), with six
point targets and complex Gaussian noise. The antenna beam limits each target's synthetic aperture to
600\,m, fully inside the frame, and the targets' echoes drift by 1.8 range bins across it, so the
correction of this drift, range cell migration correction (RCMC), matters.

\textit{Processing and validation.} Every pipeline computes the same steps: range compression (FFT,
matched filter and inverse FFT along each pulse); an azimuth FFT along each range bin; RCMC with an
8-tap interpolator; the azimuth filter $\exp(+j4\pi R_0(D-1)/\lambda)$, with
$D=\sqrt{1-(\lambda f_a/2v)^2}$; and an azimuth inverse FFT. A CPU reference (vDSP FFTs,
double-precision RCMC and phase) focuses every target at its exact position with textbook quality: its
impulse response is 1.06 range bins and 8.81 azimuth bins wide (theory 1.06 and 8.86), with peak
sidelobes at $-13.2$ and $-13.5$\,dB (theory $-13.26$). Every GPU and hybrid image reproduces these values
to 0.01\,dB, and the FP32 GPU image differs from the reference by $2.9\times10^{-6}$ (relative error).
Reaching this agreement took four numerical fixes, each worth knowing:
\begin{enumerate}
\item Use the baseband azimuth filter, with $(D-1)$ instead of $D$. The textbook
  $\exp(+j4\pi R_0D/\lambda)$ adds $4\pi\delta_r/\lambda\approx65$\,rad per range bin (bin spacing
  $\delta_r$), which shifts and wraps the 83\%-wide range spectrum. That is harmless for the image
  magnitude, but it breaks any band-limited interpolation of the image.
\item The remaining phase reaches $10^4$\,rad at $|f_a|=\mathrm{PRF}/2$, so we compute it in cycles, with
  a split that makes the FP32 product exact.
\item Blend the interpolator's table entries linearly. Nearest-entry lookup is discontinuous, and FP32
  and double-precision positions pick different entries for about 2\% of samples.
\item Take the fractional shift from the shift $\Delta n$ alone, not from $n+\Delta n$.
\end{enumerate}

\begin{table}[t]
\centering
\caption{Range--Doppler SAR on a $4096\times4096$ Scene, ms (Medians of Three Recorded Runs, Each a
Median of 7). GPU: GPU Time; Hybrid: Wall Clock Including CPU--GPU Synchronization; Floor: Minimum Time
for the Main-Memory Traffic at 150\,GB/s}
\label{tab:sar}
\footnotesize
\setlength{\tabcolsep}{3pt}
\begin{tabular}{@{}lrrrrr@{}}
\toprule
\textbf{Pipeline} & \textbf{Range} & \textbf{Az.\ FFT} & \textbf{RCMC+IFFT} & \textbf{Total} & \textbf{Floor} \\
\midrule
CPU reference (vDSP) & 23.6 & 21.5 & 66.1 & 113 & -- \\
GPU FP32, 4 passes & 1.89 & 3.86 & 3.05 & 8.87 & 7.2 \\
GPU FP32, 3 passes & 1.89 & 3.18 & 3.05 & \textbf{8.13} & -- \\
GPU FP16, 4 passes & 1.45 & 1.98 & 2.55 & \textbf{6.02} & 4.5 \\
GPU + SME (hybrid) & 2.54 & 9.76 & 2.60 + 9.83 & 24.9 & -- \\
\bottomrule
\end{tabular}
\end{table}

\textit{Performance.} The GPU pipeline makes four passes through main memory: fused range compression;
a transpose; the azimuth FFT along rows; and a fused kernel that gathers eight neighboring rows for
RCMC, applies the azimuth filter and runs the inverse azimuth FFT. It takes 8.87\,ms, while moving its
data through main memory takes at least 7.2\,ms, so it runs at 81\% of that limit (Table~\ref{tab:sar}).
Replacing the transpose and the row FFT with one FFT that reads the columns directly removes a pass. Its
4-byte strided reads look wasteful, but threadgroups working on neighboring columns run at the same time
and share cache lines in L2. The azimuth FFT then takes 3.18 instead of 3.86\,ms, and the whole image
8.13\,ms, $14\times$ faster than the CPU reference. Storing the three intermediate images in FP16
(Section~\ref{sec:fp16}) cuts the traffic from 64 to 40 bytes per point and the time to 6.02\,ms:
$19\times$ the reference and 74\% of its limit, with an image error of $-71$\,dB and unchanged focus
quality. FP16's limited range is handled by the fixed-shift block-floating-point schedule
of~\cite{bergach2026fp16}: range compression includes its $1/N_r$ scale, and the azimuth FFT output is
scaled by $1/N_a=1/4096$. Every intermediate value is then bounded by about the input's full scale times
the pulse length (1200 samples here), far below the FP16 maximum, whatever the scene. The fused RCMC
kernel is the slowest pass in every variant: it moves about 88\,GB/s of useful data, against about
140\,GB/s for the other passes. For context, the fused M1 pipeline of~\cite{bergach2026sar} formed a
$4096\times4096$ image in 370\,ms; algorithm details, kernels and hardware differ, so the two numbers
are not directly comparable.

\textit{The CPU matrix unit in the pipeline.} A hybrid pipeline runs the two column transforms on the
CPU's matrix unit and needs no transpose. It is nevertheless $3\times$ slower than the best GPU
pipeline: one SME column pass (9.8\,ms, about 100\GF{}) costs as much as five GPU passes. Running both
engines at once does not help either. With the GPU processing one scene and the CPU unit running column
FFTs on another, the GPU slows from 112 to 101 scenes/s and the CPU unit from 90--104 to 37--40
transforms/s. Total main-memory traffic stays at 118--120\,GB/s, the same as the GPU alone: the CPU unit
only takes its share of a fixed bandwidth. On a chip whose FFT work is limited by shared memory
bandwidth, a second compute engine adds nothing. The SME path pays only when the CPU works alone: it
computes the azimuth FFT in 9.8\,ms, against 21.7\,ms for vDSP with a transpose.

\section{Discussion}\label{sec:discussion}

\textit{The on-chip rule on M6.} The 2015 rule (compute the largest block that fits on chip) still
decides how to write the \emph{kernel}, with two amendments. First, ``on chip'' means a threadgroup's
registers plus its threadgroup memory, not threadgroup memory alone; this moves the one-pass limit from
4096 to at least 16384 points. Second, a batch-level rule decides the speed: which memory level holds
the batch (L2 up to about 2\,MiB, SLC up to about 16\,MiB, main memory beyond), with each extra pass
through main memory dividing the limit.

\textit{Matrix units.} One bound decides both matrix units: a dense radix-$r$ block pays only if the unit
is more than $8r/(5\log_2r)$ times faster than the vector path. The GPU's unit reaches that bound at
radix 8 only at peak speed and falls short in practice; the CPU's exceeds it by an order of magnitude.
Neither adds throughput once the data lives in main memory, where the whole chip shares one bandwidth.
Energy measurements point the same way: most of a computer's energy goes into keeping and moving data,
not into arithmetic~\cite{bergach2026energy}.

\textit{Limitations.} All results come from one M6 machine (12-core GPU, 32\,GiB) on one OS build, on AC
power, without power or clock measurements. MLX is timed by wall clock. Our GPU kernels for $N\le4096$
are the M1-era kernels of~\cite{bergach2026fft}, not re-tuned for M6. We tried one matrix-unit FFT
design (radix 16, one threadgroup per transform); int8 or bfloat16 operands, and larger batched designs
that raise $K$, remain untested. The SME kernel supports only $N=8^s$ in FP32 and does not fold the
layout conversion into its first and last stages. The radar scene is simulated (point targets, flat
earth, zero squint) at a single size, without secondary range compression, and the CPU reference checks
correctness; it is not a tuned CPU pipeline.

\section{Conclusion}

On Apple M6, the FFT is a data-movement problem at every size we measured. Large batches run at the
main-memory limit (our kernels and MPSGraph within 3\%), and cached batches are limited by the SLC or L2.
Only two decisions still matter: how many times the data crosses main memory, and how many bytes each
value takes. Keeping 8192- and 16384-point transforms in registers saves one pass ($2.2\times$ and
$1.85\times$ MPSGraph), and FP16 storage halves the bytes ($4.4\times$ at $N=8192$). More arithmetic,
even the GPU's 19.6\,TFLOPS matrix units, does not help. The CPU's matrix unit is the exception: at about
$47\times$ the speed of its vector units, it lets a radix-8 kernel beat vDSP by $1.5\times$ on batches
and by up to $5\times$ on matrix columns. Next to the GPU, however, it adds no throughput. A complete
$4096\times4096$ radar image takes 8.1\,ms in FP32, or 6.0\,ms with FP16 intermediates (74\% of its
main-memory limit), on the M6 GPU.

\smallskip
\noindent\textbf{Reproducibility.} All Metal kernels, the SME2 FFT, the SAR simulator and pipelines,
the benchmark programs, raw logs and plotting scripts are available under the MIT license in the
\texttt{m6/} directory of \url{https://github.com/aminems/AppleSiliconFFT}.

\IEEEtriggeratref{17}
\bibliographystyle{IEEEtran}
\bibliography{references}

@IEEEtranBSTCTL{IEEEexample:BSTcontrol,
  CTLdash_repeated_names = "no",
}

@phdthesis{bergach2015thesis,
  author       = {Mohamed Amine Bergach},
  title        = {Adaptation du calcul de la Transform\'{e}e de {Fourier} Rapide
                  sur une architecture mixte {CPU}/{GPU} int\'{e}gr\'{e}e},
  school       = {Universit\'{e} Nice Sophia Antipolis},
  year         = {2015},
  note         = {{In} {French}},
  url          = {https://theses.hal.science/tel-01245958},
}

@article{bergach2015conference,
  author       = {Mohamed Amine Bergach and Emilien Kofman and Robert de Simone
                  and Serge Tissot and Michel Syska},
  title        = {Efficient {FFT} mapping on {GPU} for radar processing
                  application: modeling and implementation},
  journal      = {arXiv preprint arXiv:1505.08067},
  year         = {2015},
  url          = {https://arxiv.org/abs/1505.08067},
}

@article{bergach2026fft,
  author       = {Mohamed Amine Bergach},
  title        = {Beating {vDSP}: A 138~{GFLOPS} Radix-8 {Stockham} {FFT} on
                  {Apple Silicon} via Two-Tier Register-Threadgroup Memory
                  Decomposition},
  journal      = {arXiv preprint arXiv:2603.27569},
  year         = {2026},
}

@book{vanloan1992fft,
  author       = {Charles F. {Van Loan}},
  title        = {Computational Frameworks for the Fast {Fourier} Transform},
  series       = {Frontiers in Applied Mathematics},
  number       = {10},
  publisher    = {SIAM},
  year         = {1992},
}

@article{frigo2005fftw,
  author       = {Matteo Frigo and Steven G. Johnson},
  title        = {The Design and Implementation of {FFTW3}},
  journal      = {Proceedings of the IEEE},
  volume       = {93},
  number       = {2},
  pages        = {216--231},
  year         = {2005},
  doi          = {10.1109/JPROC.2004.840301},
}

@inproceedings{govindaraju2008fft,
  author       = {Naga K. Govindaraju and Brandon Lloyd and Yuri Dotsenko
                  and Burton Smith and John Manferdelli},
  title        = {High Performance Discrete {Fourier} Transforms on Graphics
                  Processors},
  booktitle    = {Proceedings of the ACM/IEEE Conference on Supercomputing (SC)},
  year         = {2008},
  doi          = {10.1109/SC.2008.5213922},
}

@article{tolmachev2023vkfft,
  author       = {Dmitrii Tolmachev},
  title        = {{VkFFT} -- A Performant, Cross-Platform and Open-Source
                  {GPU} {FFT} Library},
  journal      = {IEEE Access},
  volume       = {11},
  pages        = {12039--12058},
  year         = {2023},
  doi          = {10.1109/ACCESS.2023.3242240},
}

@misc{apple_vdsp_fft,
  author       = {{Apple Inc.}},
  title        = {{vDSP} Fast {Fourier} Transforms ({Accelerate} Framework)},
  howpublished = {Apple Developer Documentation},
  note         = {{Accessed} Sep.\ 2026},
  url          = {https://developer.apple.com/documentation/accelerate/vdsp/fft},
}

@misc{corsix_amx,
  author       = {Peter Cawley},
  title        = {{Apple} {AMX} Instruction Set},
  howpublished = {GitHub repository},
  note         = {{Accessed} Sep.\ 2026},
  url          = {https://github.com/corsix/amx},
}

@article{hubner2025apple,
  author       = {Paul H{\"u}bner and Andong Hu and Ivy Peng and Stefano Markidis},
  title        = {Apple vs.\ Oranges: Evaluating the {Apple Silicon} {M}-Series
                  {SoCs} for {HPC} Performance and Efficiency},
  journal      = {arXiv preprint arXiv:2502.05317},
  year         = {2025},
}

@inproceedings{li2021tcfft,
  author       = {Binrui Li and Shenggan Cheng and James Lin},
  title        = {{tcFFT}: A Fast Half-Precision {FFT} Library for {NVIDIA}
                  {Tensor Cores}},
  booktitle    = {IEEE International Conference on Cluster Computing (CLUSTER)},
  year         = {2021},
}

@manual{nvidia_cufft,
  author       = {{NVIDIA Corporation}},
  title        = {{cuFFT} Library User's Guide},
  year         = {2024},
  url          = {https://docs.nvidia.com/cuda/cufft/index.html},
}

@manual{nvidia_cufftdx,
  author       = {{NVIDIA Corporation}},
  title        = {{cuFFTDx} -- {CUDA} {FFT} Device Extensions},
  year         = {2020},
  url          = {https://docs.nvidia.com/cuda/cufftdx/index.html},
}

@book{cumming2005sar,
  author       = {Ian G. Cumming and Frank H. Wong},
  title        = {Digital Processing of Synthetic Aperture Radar Data:
                  Algorithms and Implementation},
  publisher    = {Artech House},
  year         = {2005},
}

@inproceedings{wu2025turbofft,
  author       = {Shixun Wu and Yujia Zhai and Jinyang Liu and Jiajun Huang
                  and Zizhe Jian and Huangliang Dai and Sheng Di
                  and Franck Cappello and Zizhong Chen},
  title        = {{TurboFFT}: Co-Designed High-Performance and Fault-Tolerant
                  Fast {Fourier} Transform on {GPUs}},
  booktitle    = {Proceedings of the 30th ACM SIGPLAN Annual Symposium on
                  Principles and Practice of Parallel Programming (PPoPP)},
  year         = {2025},
  doi          = {10.1145/3710848.3710853},
}

@article{bergach2026sar,
  author       = {Mohamed Amine Bergach},
  title        = {From 8 Seconds to 370~ms: Kernel-Fused {SAR} Imaging on
                  {Apple Silicon} via Single-Dispatch {FFT} Pipelines},
  journal      = {arXiv preprint arXiv:2604.03585},
  year         = {2026},
}

@article{bergach2026fp16,
  author       = {Mohamed Amine Bergach},
  title        = {Range, Not Precision: Block-Floating-Point Half-Precision {FFT}
                  and {SAR} Imaging on {Apple Silicon}},
  journal      = {arXiv preprint arXiv:2605.28451},
  year         = {2026},
}

@inproceedings{hong1981io,
  author       = {Jia-Wei Hong and H. T. Kung},
  title        = {{I/O} Complexity: The Red-Blue Pebble Game},
  booktitle    = {Proceedings of the 13th Annual ACM Symposium on Theory of Computing (STOC)},
  pages        = {326--333},
  year         = {1981},
}

@article{williams2009roofline,
  author       = {Samuel Williams and Andrew Waterman and David Patterson},
  title        = {Roofline: An Insightful Visual Performance Model for Multicore Architectures},
  journal      = {Communications of the ACM},
  volume       = {52},
  number       = {4},
  pages        = {65--76},
  year         = {2009},
}

@misc{apple_mpsgraph,
  author       = {{Apple Inc.}},
  title        = {{MPSGraph} Fast {Fourier} Transform ({Metal Performance
                  Shaders Graph})},
  howpublished = {Apple Developer Documentation},
  note         = {{Accessed} Sep.\ 2026},
  url          = {https://developer.apple.com/documentation/metalperformanceshadersgraph/mpsgraph},
}

@misc{mlx2023,
  author       = {Awni Hannun and Jagrit Digani and Angelos Katharopoulos and
                  Ronan Collobert},
  title        = {{MLX}: Efficient and Flexible Machine Learning on {Apple Silicon}},
  howpublished = {GitHub repository},
  note         = {{Version} 0.29.3},
  url          = {https://github.com/ml-explore/mlx},
  year         = {2023},
}

@manual{arm_sme,
  author       = {{Arm Ltd.}},
  title        = {Arm Architecture Reference Manual for {A}-profile Architecture},
  note         = {{Document} {DDI} 0487. Accessed Sep.\ 2026},
  url          = {https://developer.arm.com/documentation/ddi0487/latest},
}

@inproceedings{bergach2014scaling,
  author       = {Mohamed Amine Bergach and Serge Tissot and Michel Syska
                  and Robert de Simone},
  title        = {Scaling Performance of {FFT} Computation on an Industrial
                  Integrated {GPU} Co-processor: Experiments with Algorithm
                  Adaptation},
  booktitle    = {DATE 2014 Friday Workshop -- 3PMCES},
  year         = {2014},
}

@article{bergach2026butterfly,
  author       = {Mohamed Amine Bergach},
  title        = {Dual-Select {FMA} Butterfly for {FFT}: Eliminating Twiddle Factor
                  Singularities with Bounded Precomputed Ratios},
  journal      = {arXiv preprint arXiv:2604.00567},
  year         = {2026},
}

@article{bergach2026shortest,
  author       = {Mohamed Amine Bergach},
  title        = {Shortest-Path {FFT}: Optimal {SIMD} Instruction Scheduling via
                  Graph Search},
  journal      = {arXiv preprint arXiv:2604.04311},
  year         = {2026},
}

@article{bergach2026energy,
  author       = {Mohamed Amine Bergach},
  title        = {The Price of Remembering: A Calibrated Energy Law for Computation},
  journal      = {arXiv preprint arXiv:2609.00744},
  year         = {2026},
}

@article{bergach2026int4,
  author       = {Mohamed Amine Bergach},
  title        = {When Quantization Is Free: An int4 {KV} Cache That Outruns fp16
                  on {Apple Silicon}},
  journal      = {arXiv preprint arXiv:2605.05699},
  year         = {2026},
}

@article{pease1968adaptation,
  author       = {Marshall C. Pease},
  title        = {An Adaptation of the Fast {Fourier} Transform for Parallel
                  Processing},
  journal      = {Journal of the ACM},
  volume       = {15},
  number       = {2},
  pages        = {252--264},
  year         = {1968},
}

@article{remke2024hellosme,
  author       = {Stefan Remke and Alexander Breuer},
  title        = {Hello {SME}! {Generating} Fast Matrix Multiplication Kernels
                  Using the {Scalable Matrix Extension}},
  journal      = {arXiv preprint arXiv:2409.18779},
  year         = {2024},
}

@misc{applesiliconfft,
  author       = {Mohamed Amine Bergach},
  title        = {{AppleSiliconFFT}: {FFT} Kernels, Benchmarks and Data for
                  {Apple Silicon}},
  howpublished = {GitHub repository, MIT license},
  url          = {https://github.com/aminems/AppleSiliconFFT},
  year         = {2026},
}

\end{document}